\documentclass{ws-mpla}

\begin{document}

\markboth{HSIEN-HAO MEI, WEI-TOU NI, SHENG-JUI CHEN, AND SHEAU-SHI PAN}
{AXION SEARCH WITH Q \& A EXPERIMENT}

\catchline{}{}{}{}{}

\title{AXION SEARCH WITH Q \& A EXPERIMENT}

\author{\footnotesize HSIEN-HAO MEI$^{\dag}$, WEI-TOU NI$^{\ddag}$, SHENG-JUI CHEN$^{\S}$, AND SHEAU-SHI PAN$^{\S}$ \\(Q \& A Collaboration)}

\address{$^{\dag,\ddag}$Center for Gravitation and Cosmology, Department of Physics,\\
National Tsing Hua University, Hsinchu, Taiwan, 30013 Republic of China\\
$^{\S}$Center for Measurement Standards, Industrial Technology Research Institute,\\
Hsinchu, Taiwan 30015 Republic of China\\
mei@phys.nthu.edu.tw$^{\dag}$, wtni@phys.nthu.edu.tw$^{\ddag}$}

\maketitle

\pub{Received (Day Month Year)}{Revised (Day Month Year)}

\begin{abstract}
Dark matter is a focused issue in galactic evolution and cosmology. Axion is a viable particle candidate for dark matter. 
Its interaction with photon is an effective way to detect it, e.g., pseudoscalar-photon interaction will generate vacuum dichroism in a magnetic field. 
Motivated to measure the QED vacuum birefringence and to detect pseudoscalar-photon interaction, 
we started to build up the Q \& A experiment (QED [Quantum Electrodynamics] and Axion experiment) in 1994. 
In this talk, we first give a brief historical account of planet hunting and dark matter evidence. 
We then review our 3.5 m Fabry-Perot interferometer together with our results of measuring vacuum dichroism and gaseous Cotton-Mouton effects. 
Our first results give (-0.2 $\pm$ 2.8) $\times$ 10$^{-13}$ rad/pass, 
at 2.3 T with 18,700 passes through a 0.6 m long magnet, for vacuum dichroism measurement. 
We are upgrading our interferometer to 7 m arm-length with a new 1.8 m 2.3 T permanent magnet capable of rotation up to 13 cycles per second. 
We aim at 10 nrad/$\sqrt{Hz}$ optical sensitivity with 532 nm cavity finesse around 100,000.
When achieved, vacuum dichroism would be measured to 8.6 $\times$ 10$^{-17}$ rad/pass in about 50 days, 
and QED birefringence would be measured to 28 \%.

\end{abstract}

\ccode{PACS Nos.: 14.80.Va, 98.89.-k, 12.20.-m, 04.80.-y, 07.60.Ly, 07.60.Fs, 33.55.+b.}

\section{Introduction}	

A focus of this workshop is on dark matter. Evidence for dark matter comes from deviations of observed astrophysical matter distribution 
from predictions of Newtonian dynamics and Newtonian gravitation. Logically, the deviations could lead either to missing mass called dark matter 
or modification of Newtonian dynamics and Newtonian gravitation. Let us first look into historical development of planet discoveries 
and Newtonian predictions.

Newtonian theory is based on planetary observations and Galileo's experiment on inclined planes. 
After Herschel's discovery of the planet Uranus in 1781, the observed Uranus trajectory persistently wandered away from its expected Newtonian path. 
In 1834, Hussey suggested that the deviation is due to perturbation of an undiscovered planet. In 1846, 
Le Verrier predicted the position of this new planet. On September 25, 1846, Galle and d'Arrest found the new planet, 
Neptune, within one degree of arc of Le Verrier's calculation.\cite{1}

With the discovery of Neptune, Newton's theory of gravitation was at its peak. 
In 1859, Le Verrier discovered the anomalous perihelion advance of Mercury.\cite{2} 
Efforts to account for the anomalous perihelion advance of Mercury went into two general directions: 
(i) searching for the planet Vulcan, intra-Mercurial matter and the like; 
(ii) modification of the gravitation law. Both kinds of efforts were not successful in the last half of the 19th century. 
The successful solution awaited the development of general relativity in 1915.\cite{3} Towards the middle of nineteenth century, 
as the orbit determination of Mercury reached 10$^{-8}$ fractional precision, 
relativistic effect of gravity showed up as the anomalous perihelion advance.\cite{4} 
Around the same time, the anomaly in lunar orbit was eventually explained by energy dissipation of earth tides caused by lunar motion.

Astronomers hoped that similar irregularities in Neptune's orbit would lead to another planet discovery. Since Neptune's orbit period is 165 years, 
there were insufficient observations and predictions of another unknown planet had to be based on perturbations in Uranus' orbit 
after subtracting Neptune's effect. Pickering made the first prediction in 1909 for a remote planet O. 
Lowell made next attempt in 1915 calling his unknown object Planet X. Working in Lowell Observatory, Tombaugh discovered Planet X, 
named shortly after as Pluto, in 1930. However, Pluto's mass seemed too small to cause the calculated perturbations of Uranus. 
The discovery that Pluto has a satellite in 1978 permitted its mass determination. 
``{\em Pluto's mass is too small; Pluto was not found because it was correctly predicted.}''\cite{5} 
Sixty-two years (in 1992) after the discovery of Pluto, another trans-Neptunian object was discovered. 
We now have over one thousand such objects forming Kuiper belt; some of them are bigger than Pluto.

Thus we see that planet orbit determination and planet hunting leads to progress in planetary science, geophysics, 
theory of gravity and discovery in the outer solar system.

In 1933, Zwicky found evidence of dark matter from the velocity dispersion of the galaxies in the Coma cluster.\cite{6,7} More recently, 
X-ray observations indicated that the temperature of cluster gas is too high according to virial theorem of Newtonian theory 
and requires a factor of 5 more matter other than visible baryonic matter. 
Galactic rotation curves\cite{8} provide evidence for galactic dark matter if one assumes Newtonian theory.  
On the cosmological scale, dark matter is important for structure formation to agree with observations.

Since the presence of dark matter is inferred from its gravitational effects on visible matter, 
until it is directly detected a viable possibility is that gravity law needs to be modified.\cite{9}

Aside from black holes and other possibilities, there are various particle candidates for non-baryonic dark matter: 
standard model neutrinos, sterile neutrinos, axions, supersymmetric candidates (neutralinos, sneutrinos, gravitinos, axinos), 
light scalar dark matter, dark matter from little Higgs models, Kaluza-Klein states, superheavy dark matter, Q-balls, mirror particles, 
CHArged Massive Particles (CHAMPs), self interacting dark matter, D-matter, cryptons, superweakly interacting dark matter, brane world dark matter, 
heavy fourth generation neutrinos, etc.\cite{10}

In this presentation, we will focus on axion search through its interaction with photons. In the next section, 
we review pseudoscalar-photon interaction and axions. In section 3, we describe the Q \& A (QED [Quantum Electrodynamics] and Axion experiment) experiment and its Phase II results. 
In section 4, we present the goal and progress of our current upgrade. In section 5, we conclude with discussion and outlook.

\vspace{-0.1cm}
\section{Pseudoscalar-photon interaction and axions}

In the 5th Patras Workshop on Axions, WIMPs and WISPs held at the University of Durham on 13-17 July 2009, 
three motivations were presented. In the bottom-up approach,\cite{11} 
axion is considered as a Goldstone boson associated with spontaneously broken PQ symmetry\cite{12} to fix the strong $CP$ problem. 
The name of axion was proposed by Wilczek as detergent AXION (AXION is a commercial brand of detergent) to clean up the strong $CP$ problem. 
As the original axions\cite{13,14} were not found, invisible axions\cite{15,16,17} were proposed. 

Top-down motivation\cite{18} comes from superstring theory. In supersymmetry/ supergravity, 
an appropriate action connects gauge and axionic couplings through a single holomorphic function. 
In type IIA/B superstring theory, axion comes from a Ramond-Ramond antisymmetrical field reduced on the cycle (compactified space). 
Axions also arise for heterotic string and M-theory. In superstring theory, 
``{\em the model-independent axion appears in every perturbative string theory, and is closely related to the graviton and dilaton.}''\cite{19}

The third motivation\cite{20,21} (historically the first approach) comes from a phenomenological study of equivalence principles. 
In 1973, we studied the relationship of Galilio Equivalence Principle (WEP I) and Einstein Equivalence Principle (EEP) in a framework 
(the $\chi-g$ framework) of electromagnetism and charged particles, and found the following example with interaction Lagrangian density
\vspace*{-0.05cm}
\begin{equation}
{\cal L}_I = - {1\over 16\pi}g^{ik}g^{jl}F_{ij}F_{kl} - {1\over 16\pi}\varphi F_{ij}F_{kl}e^{ijkl} - A_kj^k\sqrt{-g} - \sum_I m_I{ds_I\over dt}\delta({\bf x}-{\bf x_I}),
\label{eq1}
\end{equation}
\vspace*{-0.05cm}
as an example which obeys WEP I, but not EEP.\cite{22,23,24} The nonmetric part of this theory is
\vspace*{-0.05cm}
\begin{equation}
{\cal L}_I^{(NM)} = - {1\over 16\pi}\sqrt{-g}\varphi\varepsilon^{ijkl}F_{ij}F_{kl} = - {1\over 4\pi}\sqrt{-g}\varphi_{,i}\varepsilon^{ijkl}A_jA_{k,l}\hbox{(mod div)},
\label{eq2}
\end{equation}
\vspace*{-0.05cm}
where `mod div' means that the two Lagrangian densities are related by partial integration in the action integral. The Maxwell equations\cite{22,24} are
\begin{equation}
F^{ik}_{\vert k} + \varepsilon^{ikml}F_{km}\varphi_{,l} = -4\pi j^i,
\label{eq3}
\end{equation}
where the derivation `$\vert$' is with respect to the Christoffel connection of $g^{ik}$. 
The Lorentz force law is the same as in metric theories of gravity or general relativity. 
auge invariance and charge conservation are guaranteed. The Maxwell equations (\ref{eq3}) are also conformally invariant. 
Axial-photon interaction induces energy level shift in atoms and molecules, and polarization rotations in electromagnetic wave propagation. 
Empirical tests of the pseudoscalar-photon interaction (\ref{eq2}) were analyzed in 1973; at that time it was only loosely constrained. 
Now we have effective constraints on polarization rotation in the electromagnetic wave propagation from astrophysical polarization observations 
and CMB polarization observations for massless or nearly massless case.\cite{21,25} Axion with mass is a viable candidate for dark matter search. 
Recently Laboratory experiments have started to give constraints on them.\cite{26}$^-$\cite{F} 

The rightist term in equation (\ref{eq2}) is reminiscent of Chern-Simons\cite{34} term $e^{\alpha\beta\gamma}A_\alpha F_{\beta\gamma}$. 
There are two differences: (i) Chern-Simons term is in 3 dimensional space; (ii) Chern-Simons term in the integral is a total divergence. 
A term similar to the one in equation (\ref{eq2}) (axion-gluon interaction) occurs in QCD in an effort 
to solve the strong $CP$ problem with the electromagnetic field replaced by gluon field.\cite{12,13,14} Carroll, Field and Jackiw\cite{35} 
proposed a modification of electrodynamics with an additional $e^{ijkl}V_iA_jF_{kl}$ term with $V_i$ a constant vector. 
This term is a special case of the term $e^{ijkl}\varphi F_{ij}F_{kl} \hbox{(mod div)}$ with $\varphi_{,i} = -{1\over2} V_i$. 
Various terms in the Lagrangians discussed here are listed in Table \ref{tb1} for comparison. 

\begin{table}[h]
\vspace{-0.3cm}
\tbl{Various terms in the Lagrangian and their meaning.}
{
\begin{tabular}{cccc} 
\hline\hline
Term & Dimension & Reference & Meaning \\
\colrule
$e^{\alpha\beta\gamma}A_\alpha F_{\beta\gamma}$ & 3 & $\begin{array}{c}\hbox{Chern-Simons\cite{34}} \\ (1974) \end{array}$ & $\begin{array}{c}\hbox{Integrand for} \\ \hbox{topological invariant} \end{array}$ \\
\vspace*{-0.1cm} & & & \\
$e^{ijkl}\varphi F_{ij}F_{kl}$ & 4 & $\begin{array}{c}\hbox{Ni\cite{22,23,24}} \\ (1973,1974,1977) \end{array}$ & $\begin{array}{c}\hbox{Pseudoscalar-photon} \\ \hbox{coupling} \end{array}$ \\
\vspace*{-0.1cm} & & & \\
$e^{ijkl}\varphi F^{QCD}_{ij}F^{QCD}_{kl}$ & 4 & 
$\begin{array}{c}
\hbox{Peccei-Quinn\cite{12}} (1977) \\
\hbox{Weinberg\cite{13}} (1978)\\
\hbox{Wilczek\cite{14}} (1978)
\end{array}$ 
& $\begin{array}{c}\hbox{Pseudoscalar-gluon} \\ \hbox{coupling} \end{array}$ \\
\vspace*{-0.1cm} & & & \\
$e^{ijkl}V_iA_jF_{kl}$ & 4 & $\begin{array}{c}\hbox{Carroll-Field-Jackiw\cite{35}} \\ (1974) \end{array}$ & $\begin{array}{c}\hbox{External constant} \\ \hbox{vector coupling} \end{array}$ \\ 
\hline\hline
\end{tabular}\label{tb1}
}
\vspace*{-0.3cm}
\end{table}

In the Peccei-Quinn-Weinberg-Wilczek models, axion-photon interaction may or may not be induced. In terms of Feynman diagram, 
the interaction (\ref{eq2}) gives a 2-photon-pseudo-scalar vertex. With this interaction, vacuum becomes birefringent and dichroic.\cite{36,37,38,39,40}

Dichroic materials have the property that their absorption constant varies with polarization. For axion models with (\ref{eq2}), 
photon interacts with magnetic field has a cross section to be converted into axion and leaks away. The vacuum with magnetic field becomes absorptive. 
Since the cross section depends on polarization, so is the absorption. When polarized light goes through vacuum with magnetic field, 
its polarization is rotated due to difference in absorption in 2 principal directions of the vacuum for the 2 polarization components. 
The polarization rotation $\varepsilon$ of the photon beam for light entering the magnetic-field region polarized at an angle of $\theta$ 
to the magnetic field is\cite{26}  
\vspace*{-0.2cm}
\begin{equation}
\varepsilon = \left({B\omega\over Mm_\varphi^2}\right)^2 \sin^2\left({m_\varphi^2L\over 4\omega}\right) \sin2\theta \buildrel {\buildrel {m_\varphi^2L\over 4\omega} \ll 1 \over \downarrow} \over \simeq \left({BL\over 4M}\right)^2 \sin2\theta,
\label{eq4}
\end{equation}
where $m_\varphi$ is mass of the axion, $M$ the axion-photon interaction energy scale, 
$\omega$ photon circular frequency and $L$ the magnetic-region length. The approximation is valid 
under the condition
\vspace*{-0.1cm}
\begin{equation}
m_\varphi^2 < {2\pi\omega\over L},
\label{eq5}
\end{equation}
to preserve the relative phase of the axion and photon fields. Since axions do not reflect at the mirrors, for multi-passes, 
the rotation effect increases by number $N$ of passes. Therefore for the case condition (\ref{eq5}) is satisfied, 
the polarization rotation effect is proportional to $NB^2L^2$. 
Axion leaking away has a cross section to interact with magnetic field in another region to regenerate photon. 
This is the theoretical theme for photon regereration experiment (also known as LSW, `Light Shining through Wall').

Current optical experiments to measure the vacuum dichroism ($\varepsilon$, as that in Eq. (\ref{eq4})), 
vacuum birefringence ($\psi$) through pseudoscalar-photon interaction or QED, 
and to detect photon regeneration are listed in Table \ref{tb2}. 
We present our Q \& A experiment in the next section.

\begin{table}[h]
\vspace*{-0.1cm}
\tbl{Current laser experiments for detecting dark matter candidates, pseudoscalar-photon interaction, or aiming at nonlinear QED effects.}
{
\begin{tabular}{llllll} 
\hline\hline
Collaboration & running & $\begin{array}{c} \hbox{results} \\ \hbox{in } \varepsilon \end{array}$ & $\begin{array}{c} \hbox{results} \\ \hbox{in } \psi \end{array}$ & $\begin{array}{c} \hbox{results} \\ \hbox{in LSW} \end{array}$ & $\begin{array}{c} \hbox{aiming} \\ \hbox{at } \psi_{QED} \end{array}$ \\
\colrule
ALPS & $\surd$ & & & $\surd$\cite{33,F} & \\
BFRT & & $\surd$\cite{26} & $\surd$\cite{26} & $\surd$\cite{26} & \\
BMV & $\surd$ & & & $\surd$\cite{29,B} & $\surd$ \\
GammeV & $\surd$ & & & $\surd$\cite{30,C} & \\
LIPSS & $\surd$ & & & $\surd$\cite{31,D} & \\
OSQAR & $\surd$ & & & $\surd$\cite{32} & \\
PVLAS & $\surd$\cite{E} & $\surd$\cite{27,A} & $\surd$\cite{A} & $\surd$\cite{A} & $\surd$ \\
Q \& A & $\surd$ & $\surd$\cite{28} & & & $\surd$ \\
\hline\hline
\vspace*{-1.1cm}
\end{tabular}\label{tb2}
}
\end{table}

\section{Q \& A experiment}

In 1991, when Tsubono from University of Tokyo visited our Gravitation Laboratory in National Tsing Hua University, 
we discussed the technical development of ultra-high sensitive interferometers for gravity-wave detection. 
During the last day before his departure, we pondered about how we could apply these developed techniques for fundamental physics 
and we discussed the possibility of doing the interferometric QED (Quantum Electrodynamics) tests. After analyzing the sensitivities, 
we believe that the QED birefringence would be measurable.\cite{41}

After the call for EOI's (Expression of Interest) of using the onsite SSC (Superconducting Super Collider) facilities in March, 1994 by DOE (Department of Energy) of USA, 
we submitted a joint EOI with a US team.\cite{42} The topic of this EOI was chosen as one of the six topics for project definition study proposals. 
We then submitted such a proposal\cite{43} in June and finished the study at the end of October, 1994.\cite{44} The project definition review was well-received. 
A five-year proposal\cite{45} was submitted to the National Science Council of the Republic of China for the ROC part of the funding simultaneously. 
This proposal was approved in January, 1995 pending on the approval of the US proposal of the collaboration. 
Partial funding was allocated for the first year. However, due to lack of potential funding of the US counterpart, 
this program of collaboration was halted.

Nevertheless, in 1994, we started to build the experimental facility for the Q \& A experiment\cite{46,47,48} acquiring two vacuum tanks of the laser-interferometric gravity-wave detector type 
and working on the measurement of mirror birefringence.\cite{49} Since 1991 we have worked on precision interferometry --- laser stabilization schemes, 
laser metrology and Fabry-Perot interferometers. With these experiences, we started in 1994 to build a 3.5 m prototype interferometer 
for measuring vacuum birefringence and improving the sensitivity of axion search as part of our continuing effort in precision interferometry. 
In June 1994, we met the PVLAS people in the Marcel Grossmann Meeting at Stanford, exchanged a few ideas and encouraged each other. 
PVLAS also started in the same year adapting their earlier scheme proposed in 1979.\cite{50}

In 2002, we finished Phase I of constructing the 3.5 m prototype interferometer and made some Cotton-Mouton coefficient 
and Verdet coefficient measurements.\cite{51} Starting 2002, we have been in Phase II of Q \& A experiment. 
The results of Phase II on dichroism and Cotton-Mouton effect measurement have been reported.\cite{28,52} 
We are starting Phase III of our Q \& A experiment of extending the 3.5 m interferometer to 7 m with upgrades. 
In following, we review our experimental setup and summarize our vacuum and gas measurement results.

\vspace*{-0.4cm}
\subsection{Experimental setup, optical sensitivity and first results}
\vspace*{-0.05cm}

\begin{figure}[ph]
\vspace*{-0.3cm}
\centerline{\psfig{file=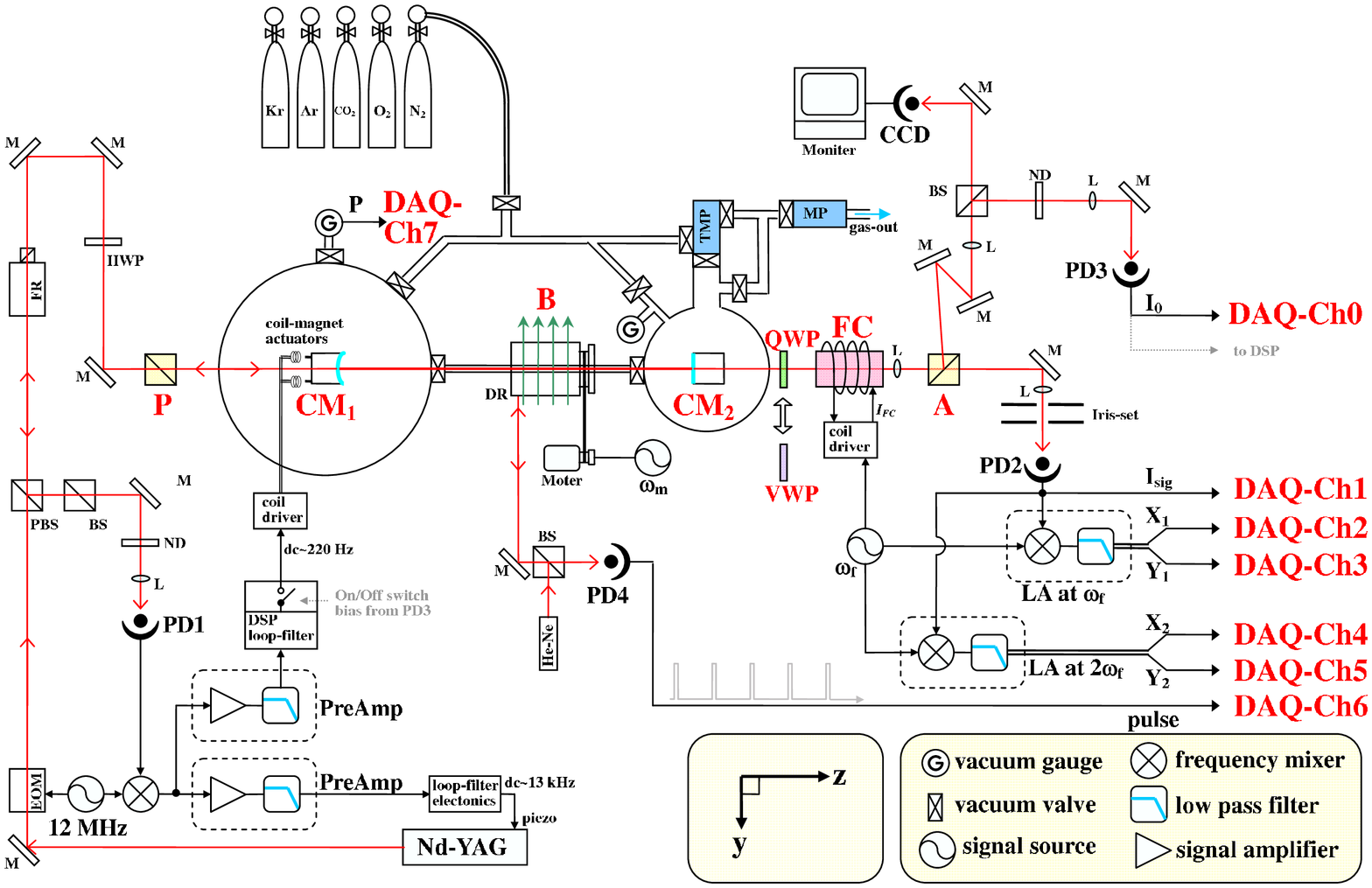,width=3.5in,angle=270}}
\caption{The experimental setup of Q \& A experiment Phase II.\protect\label{fg1}}
\vspace*{-0.6cm}
\end{figure}

The schematic of the setup of Phase II is shown in Fig. \ref{fg1}. 
The 3.5 m prototype interferometer is formed using a high-finesse Fabry-Perot interferometer together with a high-precision ellipsometer. 
The two high-reflectivity mirrors of the 3.5 m prototype interferometer are suspended separately from two X-pendulum-double pendulum suspensions 
mounted on two isolated tables fixed to ground using bellows inside two vacuum chambers. The sub-systems are described in references.\cite{28,52,53,54,55} 
Our results in this phase give $(-0.2 \pm 2.8) \times 10^{-13}$ rad/pass with 18,700 passes through a 2.3 T 0.6 m long magnet 
for vacuum dichroism measurement, constraining pseudo-scalar-photon interaction to $M > 6 \times 10^5$ GeV for $m_\varphi < 1.7$ meV, 
and limit milli-(mini-)charged fermions meaningfully.\cite{28}
We have performed a 78 \% duty cycle within 30 days with over-all optical sensitivity of the apparatus around 10$^{-6}$ rad/$\sqrt{\hbox{Hz}}$ in dichroism 
and 10$^{-6}/\sqrt{\hbox{Hz}}$ in ellipticity.

\vspace*{-0.4cm}
\subsection{Calibration of the apparatus through gaseous birefringence}
\vspace*{-0.05cm}

Upon passing through a medium with transverse magnetic field, linearly polarized light becomes elliptical polarized. 
Cotton and Mouton first investigated this in detail in 1905, and the phenomenon is known as Cotton-Mouton effect (CME). 
We use our Q \& A apparatus to measure the CMEs at wavelength 1064 nm in N$_2$, O$_2$, CO$_2$, Ar, and Kr
in a magnetic field $B$ = 2.3 T at pressure $P$ = 0.5-300 Torr and temperature $T$ = 295-298 K. 
Our measured results\cite{52} are compiled in Table \ref{tb3} and used to calibrate the apparatus for vacuum results. 
For the Cotton-Mouton coefficient, we follow the common convention\cite{56} of the normalized Cotton-Mouton birefringence $\Delta n_u$ at $P$ = 1 atm 
and $B$ = 1 T. Our results agree with the PVLAS results\cite{57,58,59} in the common cases (Kr, N$_2$, O$_2$) within 1.2 $\sigma$. 
For Ar and CO$_2$ at 1064 nm, our results are new.

\begin{table}[h]
\vspace*{-0.1cm}
\tbl{Measured gaseous Cotton-Mouton coefficients.\protect\cite{52}}
{
\begin{tabular}{lc} 
\hline\hline
Gas & $\begin{array}{c} \hbox{Normalized Cotton-Mouton birefringence\cite{56}} \\ \Delta n_u \hbox{ at } P \hbox{ = 1 atm and } B \hbox{ = 1 T} \end{array}$ \\
\colrule
N$_2$ & ($-$ 2.02 $\pm$  0.16$^\S \pm$    0.08$^\P$) $\times$  10$^{-13}$ \\
O$_2$ & ( $-$1.79  $\pm$ 0.34$^\S \pm$  0.08$^\P$)   $\times$ 10$^{-12}$ \\
CO$_2$ &   ($-$ 4.22 $\pm$  0.27$^\S \pm$ 0.16$^\P$) $\times$   10$^{-13}$ \\
 Ar  & (4.31 $\pm$  0.34$^\S \pm$  0.17$^\P$)  $\times$  10$^{-15}$  \\
 Kr  & (8.28 $\pm$  1.26$^\S \pm$  0.32$^\P$)   $\times$ 10$^{-15}$  \\ 
\hline\hline
$\S$: Statistical uncertainty\\
$\P$: Systematic uncertainty
\end{tabular}\label{tb3}
}
\vspace*{-0.35cm}
\end{table}

\vspace*{-0.35cm}
\section{Current upgrade}
\vspace*{-0.05cm}

\begin{figure}[ph]
\vspace*{-0.4cm}
\centerline{
\begin{minipage}[h]{7.2cm}
\psfig{file=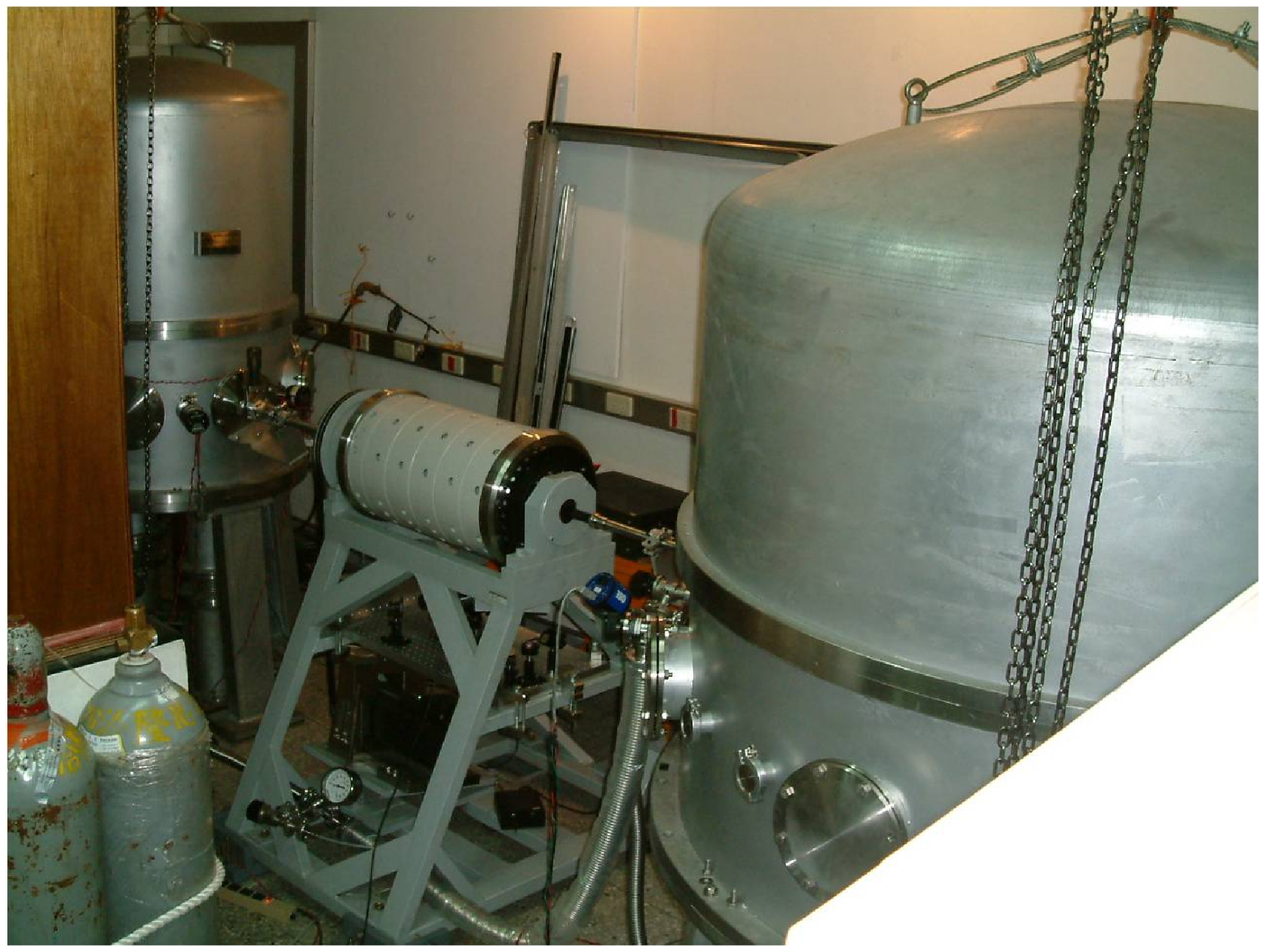,scale=0.42}
\vspace*{-0.8cm}
\caption{Q \& A Phase II.\protect\label{fg2}}
\end{minipage}
\begin{minipage}[h]{4cm}
\psfig{file=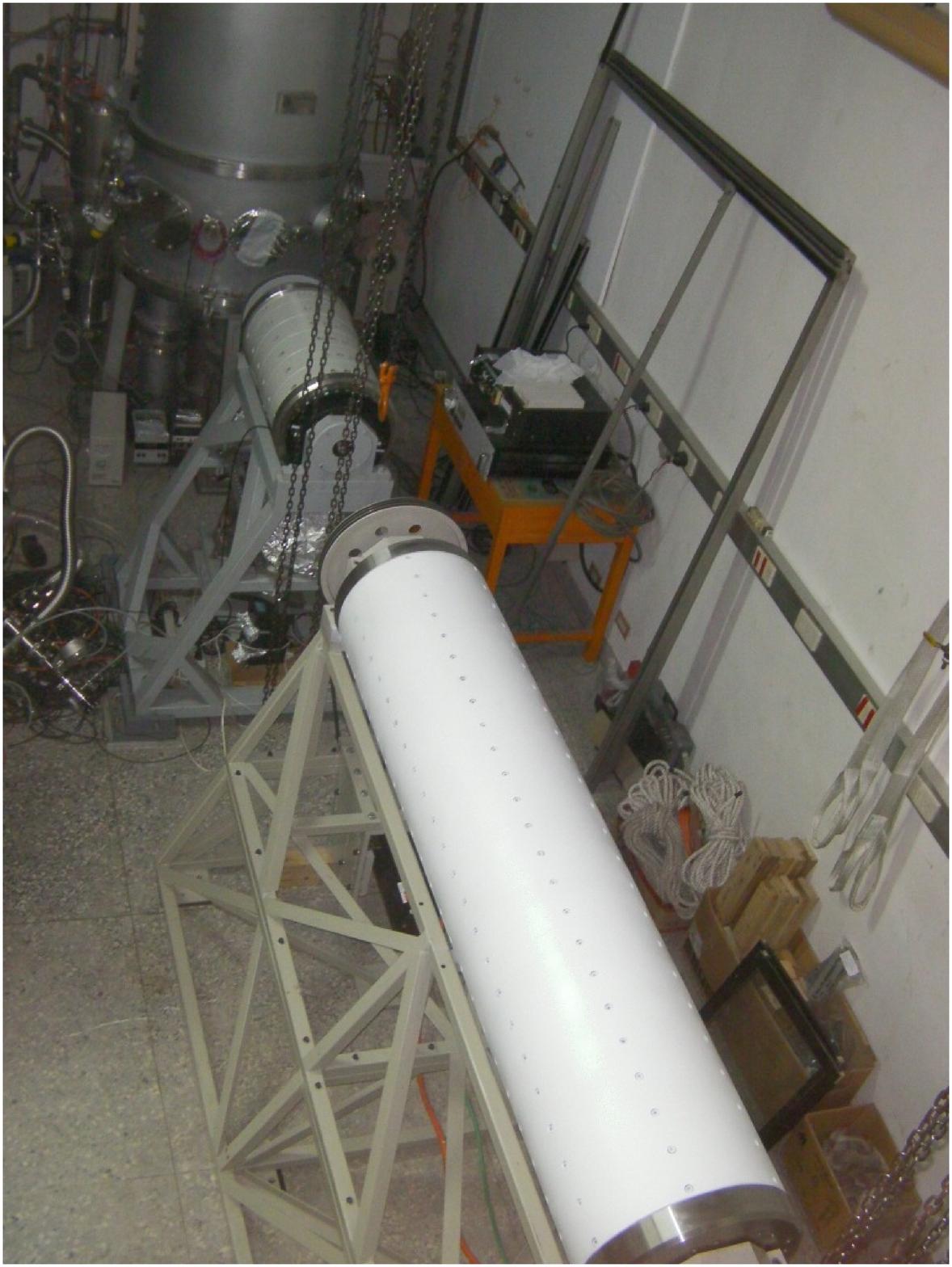,scale=0.15}
\vspace*{-0.8cm}
\caption{Q \& A Phase III.\protect\label{fg3}}
\end{minipage}
}
\vspace*{-1cm}
\end{figure}

We are currently upgrading our interferometer from 3.5 m arm-length to 7 m arm-length in Phase III. 
We have installed a new 1.8 m 2.3 T permanent magnet capable of rotation up to 13 cycles per second to enhance the physical effects. 
Fig. \ref{fg2} and Fig. \ref{fg3} show pictures of Phase II, and Phase III with magnets. 
We are working with 532 nm Nd:YAG laser as light source with cavity finesse around 100,000, and aim at 10 nrad/$\sqrt{\hbox{Hz}}$ optical sensitivity. 
With all these achieved and the upgrading of vacuum, in 50 days (with duty cycle around 78 \% as performed before) the vacuum dichroism measurement would be improved in precision 
by 3-4 orders of magnitude for each passage, which would further constrain the energy scale $M$ in Eq. (\ref{eq4}) by 2 orders;
and QED birefringence would be measured to 28 \%. 
To enhance the physical effects further, another 1.8 m magnet will be added in the future.

\section{Discussion and outlook}
\vspace*{-0.05cm}

We have heard a suite of motivations to search for dark matter and (pseudo)scalar-photon interactions. For vacuum dichroism, 
after we have improved the precision by 3-4 orders of magnitude per passage after Phase III, we should still improve it further. 
For QED birefringence, the next stage after detection is to measure the next-order effects which include hadron 
and potential new physical contribution.\cite{47,48} 
This would be possible by extending the interferometer further with more rotatable permanent magnets. 
Many useful techniques have been developed in the Gravitational Wave Detection Community. We have advocated using relevant techniques.\cite{41} 
Further progress in this experimental field is expected in the near future.

\vspace*{-0.2cm}
\section*{Acknowledgements}
\vspace*{-0.05cm}

We thank the National Science Council for supporting the Q \& A program.


\begin{thebibliography}{0}
\bibitem{1} P. Moore, {\it The Story of Astronomy}, 5th revised edition (New York, Grosset \& Dunlap Publishing, 1977). 
\bibitem{2} U. J. J. Le Verrier, ``Theorie du mouvement de Mercure'', {\it Ann. Observ. imp. Paris (M\'em.)} {\bf 5}, 1-196 (1859).
\bibitem{3} A. Einstein, {\it Preuss. Akad. Wiss. Berlin, Sitzber.} 778, 799, 831, 844 (1915).
\bibitem{4} W.-T. Ni, {\it Int. J. Mod. Phys. D} {\bf 13}, 901 (2005).
\bibitem{5} K. R. Lang, {\it The Cambridge Guide to the Solar System} (Cambridge, Univ. Press, 2003).
\bibitem{6} F. Zwicky, {\it Helv. Physica Acta} {\bf 6}, 110 (1933).
\bibitem{7} F. Zwicky, {\it Astrophys. J.} {\bf 86}, 217 (1937).
\bibitem{8} V. Rubin, N. Thonnard, W. K. Ford Jr, {\it Astrophys. J.} {\bf 238}, 471(1980); and ref.s therein.
\bibitem{9} e.g., M. Milgrom, {\it Astrophys. J.} {\bf 270}, 365 (1983).
\bibitem{10} G. Bertone, D. Hooper and J. Silk, {\it Phys. Rep.} {\bf 405}, 279 (2005).
\bibitem{11} M. Ahlers, ``Axions and other (Super-)WISPs'', arXiv:0910.2211 (2009).
\bibitem{12} R. D. Peccei and H. R. Quinn, {\it Phys. Rev. Lett.} {\bf 38}, 1440 (1977).
\bibitem{13} S. Weinberg, {\it Phys. Rev. Lett.} {\bf 40}, 233 (1978).
\bibitem{14} F. Wilczek, {\it Phys. Rev. Lett.} {\bf 40}, 279 (1978).
\bibitem{15} J. Kim, {\it Phys. Rev. Lett.} {\bf 43}, 103 (1979).
\bibitem{16} M. Dine {\it et al.}, {\it Phys. Lett.} {\bf 104B}, 1999 (1981).
\bibitem{17} M. Shifman {\it et al.}, {\it Nucl. Phys. B} {\bf 166}, 493 (1980).
\bibitem{18} J. Conlon, ``Axions, WIMPs and WISPs: Top-Down Motivation'', in Proceedings of the {\it 5th Patras Workshop on Axions, WIMPs and WISPs held at the University of Durham, 2009}, http://axion-wimp.desy.de/e30/e52240/e54379/JoeConlon.pdf (2009).
\bibitem{19} See, e.g., J. Polchinski, {\it String Theory}, volume 1 \& 2 (Cambridge 1998)
\bibitem{20} H.-H. Mei, W.-T. Ni, S.-J. Chen, and S.-s. Pan (Q \& A Collaboration), ``The status and prospects of the Q \& A experiment with some applications'', arXiv:0911.4776 (2009). 
\bibitem{21} W.-T. Ni, ``Constraints on Pseudoscalar-Photon Interaction from CMB Polarization Observations'', arXiv:0910.4317 (2009).
\bibitem{22} W.-T. Ni, ``A Nonmetric Theory of Gravity'', {\it preprint}, Montana State University, Bozeman, Montana, USA (1973). The paper is available via http://cgc.pmo.ac.cn.
\bibitem{23} W.-T. Ni, {\it Bull. Am. Phys. Soc} {\bf 19}, 655 (1974).
\bibitem{24} W.-T. Ni, {\it Phys. Rev. Lett.} {\bf 38}, 301 (1977).
\bibitem{25} See, e.g., W.-T. Ni, {\it Prog. Theor. Phys. Suppl.} {\bf 172}, 49 (2008).
\bibitem{26} R. Cameron {\it et al.}, {\it Phys. Rev. D} {\bf 47}, 3707 (1993).
\bibitem{27} E. Zavattini {\it et al.} (PVLAS Collaboration), {\it Phys. Rev. Lett.} {\bf 96}, 110406 (2006).
\bibitem{A} E. Zavattini {\it et al.} (PVLAS Collaboration), {\it Phys. Rev. D} {\bf 77} 032006 (2008).
\bibitem{E} F. Della Valle {\it et al.} (PVLAS Collaboration), arXiv:0907.2642 (2009).
\bibitem{28} S.-J. Chen, H.-H. Mei and W.-T. Ni (Q \& A Collaboration), {\it Mod. Phys. Lett. A} {\bf 22} 2815 (2007) [arXiv:hep-ex/0611050].
\bibitem{29} C. Robilliard {\it et al.} (BMV Collaboration), {\it Phys. Rev. Lett.} {\bf 99} 190403 (2007).
\bibitem{B} M. Fouch\'e {\it et al.} (BMV Collaboration), {\it Phys. Rev. D} {\bf 78} 032013 (2008). 
\bibitem{30} A. S. Chou {\it et al.} (GammeV Collaboration), {\it Phys. Rev. Lett.} {\bf 100} 080402 (2008).
\bibitem{C} A. S. Chou {\it et al.} (GammeV Collaboration), {\it Phys. Rev. Lett.} {\bf 102} 030402 (2009).
\bibitem{31} A. Afanasev {\it et al.} (LIPSS Collaboration), {\it Phys. Rev. Lett.} {\bf 101} 120401 (2008).
\bibitem{D} A. Afanasev {\it et al.} (LIPSS Collaboration), {\it Phys. Lett. B} {\bf 679} 317 (2009).
\bibitem{32} P. Pugnat {\it et al.} (OSQAR Collaboration), {\it Phys. Rev. D} {\bf 78} 092003 (2008).
\bibitem{33} K. Ehret {\it el al.} (ALPS Collaboration), {\it Nucl. Instrum. Meth. A} {\bf 612} 83 (2009).
\bibitem{F} A. Lindner (ALPS Collaboration), ``Recent Results from ALPS and Future Perspectives'', presented in the {\it AXIONS 2010 held at Univ. of Florida, Jan 15-17 2010}, http://www.phys.ufl.edu/~tanner/Axions2010/alps-lindner.pdf (2010).
\bibitem{34} S.-S. Chern and J. Simons, {\it The Annals of Mathematics (2$^{nd}$ Ser.)} {\bf 99} 48 (1974).
\bibitem{35} S. M. Carroll, G. B. Field and R. Jackiw, {\it Phys. Rev. D} {\bf 41}, 1231 (1990).
\bibitem{36} P. Sikivie, {\it Phys. Rev. Lett.} {\bf 51}, 1415 (1983). 
\bibitem{37} A. A. Anselm, {\it Yad. Fiz.} {\bf 42}, 1480 (1985). 
\bibitem{38} M. Gasperini, {\it Phys. Rev. Lett.} {\bf 59}, 396 (1987).
\bibitem{39} L. Maiani, R. Petronzio and E. Zavattini, {\it Phys. Lett. B} {\bf 175}, 359 (1986).
\bibitem{40} G. Raffelt and L. Stodolsky, {\it Phys. Rev. D} {\bf 37}, 1237 (1988).
\bibitem{41} W.-T. Ni {\it et al.}, {\it Mod. Phys. Lett. A} {\bf 6} 3671 (1991). 
\bibitem{42} W.-T. Ni {\it et al.}, ``Light Retardation in a High Magnetic Field and Search for Light Scalar/Pseudo-Scalar Particles Using Ultra-Sensitive Interferometry,'' Joint EOI (Expression of Interest) submitted to the National Science Council of the Republic of China and the Department of Energy of the United States of Amenica (April, 1994).
\bibitem{43} W.-T. Ni {\it et al.}, ``Definition Studies for a proposal to Measure the Velocity of Light in a Magnetic Field'', proposal submitted to DOE (Department of Energy) of USA (June 1994); ''Experimental Search for Spin-Coupling Interactions and Light Scalar/Pseudo-scalar Particles'', Prototype Study Proposal submitted to the National Science Council of the Republic of China (June 1994).
\bibitem{44} W.-T. Ni {\it et al.}, ``Light Retardation-Absorption and Axion (LIRA) Experiment --- A Project Definition Study for On-Site Use of the Superconducting Super Collider Assets and Facilities'', U. S. Department of Energy, R-912, Final Report, Revision 1. (November 1994).
\bibitem{45} W.-T. Ni {\it et al.}, ``Test of Quantum Electro-dynamical Birefringence and Search for Light Scalar/Pseudoscalar Particles'', A Five-Year Proposal submitted to National Science Council of the Republic of China (September, 1994).
\bibitem{46} W.-T. Ni, S.-K. King, H.-W. Cheng, J.-T. Shy, N. Mio, K. Tsubono and T. C. P. Chui, in Proceedings of the {\it Sixth Marcel Grossmann Meeting on General Relativity, held at Stanford, California, 1994} (World Scientific, Singapore, 1996).
\bibitem{47} W.-T. Ni, {\it Chin. J. Phys.} {\bf 34} 962 (1996).
\bibitem{48} W.-T. Ni, {\it Frontier Test of QED and Physics of the Vacuum}, ed. E. Zavattini {\it et al.} (Sofia: Heron Press, 1998), p. 83.
\bibitem{49} H.-W. Cheng, W.-T. Ni, N. Mio, K. Tsubono, K. Kawabe, J.-T. Shy, J.-S. Wu and T.-T. Liu, in Proceedings of the {\it International Workshop on Gravitation and Cosmology, held at National Tsing Hua Univ. on December 14-17 1995} (National Tsing Hua Univ., Hsinchu, 1995), p. 190.
\bibitem{50} E. Iacopini and E. Zavattini, {\it Phys. Lett.} {\bf 85B} 151 (1971).
\bibitem{51} J.-S. Wu, S.-J. Chen and W.-T. Ni (Q \& A Collaboration), {\it Class. Quantum Grav.} {\bf 21} S1259 (2004).
\bibitem{52} H.-H. Mei, W.-T. Ni, S.-J. Chen, and S.-s. Pan (Q \& A Collaboration), {\it Chem. Phys. Lett.} {\bf 471} 216 (2009).
\bibitem{53} S.-J. Chen, H.-H. Mei and W.-T. Ni (Q \& A Collaboration), ''Improving ellipticity detection sensitivity for the Q \& A vacuum birefringence experiment'', hep-ex/0308071 (2003).
\bibitem{54} S.-J. Chen, H.-H. Mei and W.-T. Ni (Q \& A Collaboration), {\it J. of Phys.: Conf. Series} {\bf 32} 244 (2006).
\bibitem{55} H.-H. Mei, S.-J. Chen and W.-T. Ni (Q \& A Collaboration), {\it J. of Phys.: Conf. Series} {\bf 32} 236 (2006).
\bibitem{56} C. Rizzo, A. Rizzo, D. M. Bishop, {\it Int. Rev. Phys. Chem.} {\bf 16} 81 (1997).
\bibitem{57} F. Brandi {\it et al.} (PVLAS Collaboration), {\it J. Opt. Soc. Am. B} {\bf 15} 1278 (1998).
\bibitem{58} M. Bregant {\it et al.} (PVLAS Collaboration), {\it Chem. Phys. Lett.} {\bf 392} 276 (2004). 
\bibitem{59} M. Bregant {\it et al.} (PVLAS Collaboration), {\it Chem. Phys. Lett.} {\bf 477} 415 (2009).

\end{thebibliography}
\end{document}